\documentclass[a4paper]{jpconf}
\usepackage{graphicx}
\usepackage{amssymb}

\begin{document}

\title{Correlation between lasing and transport properties in a quantum dot-resonator system}

\author{Pei-Qing Jin$^1$, Michael Marthaler$^1$, Jared H.~Cole$^{1,2}$,
Maximilian K\"{o}pke$^3$, J\"{u}rgen Weis$^3$,
Alexander Shnirman$^{4,5}$,  and Gerd Sch\"on$^{1,5}$}

\address{$^1$ Institut f\"ur Theoretische Festk\"orperphysik,
      Karlsruhe Institute of Technology, 76128 Karlsruhe, Germany}
\address{$^2$ Applied Physics, School of Applied Sciences,
      RMIT University, Melbourne 3001, Australia}
\address{$^3$ Max-Planck-Institut f\"ur Festk\"orperforschung,
      D-70569 Stuttgart, Germany}
\address{$^4$ Institut f\"{u}r Theorie der Kondensierten Materie,
      Karlsruhe Institute of Technology, 76128 Karlsruhe, Germany}
\address{$^5$ DFG Center for Functional Nanostructures (CFN),
      Karlsruhe Institute of Technology, 76128 Karlsruhe, Germany}

\ead{pei-quing.jin@kit.edu}

\begin{abstract}
We study a double quantum dot system coherently coupled to an
electromagnetic resonator. By suitably biasing the system, a
population inversion can be created between the dot levels. The
resulting lasing state exists within a narrow resonance window,
where the transport current correlates with the lasing state.
It allows probing the lasing state via a current measurement. Moreover,
the resulting narrow current peak opens perspective for applications
of the setup for high resolution measurements.
\end{abstract}

Circuit quantum electrodynamics (CQED) setups with superconducting qubits
coupled to a superconducting resonator provide possibilities to
explore quantum optics effects in new parameter regimes \cite{Schoelkopf}.
Recently lasing and cooling effects were demonstrated in such systems \cite{Astafiev,Grajcar,Marthaler}.
The strong coupling regime achieved in these setups revealed qualitatively novel phenomena.
An example is the non-monotonous behavior of the linewidth of the emission spectrum in a single qubit maser,
which is influenced by quantum noise in a characteristic way \cite{SQL1,SQL2}.

\begin{figure}[t]
\centering
\includegraphics[width=0.33\textwidth]{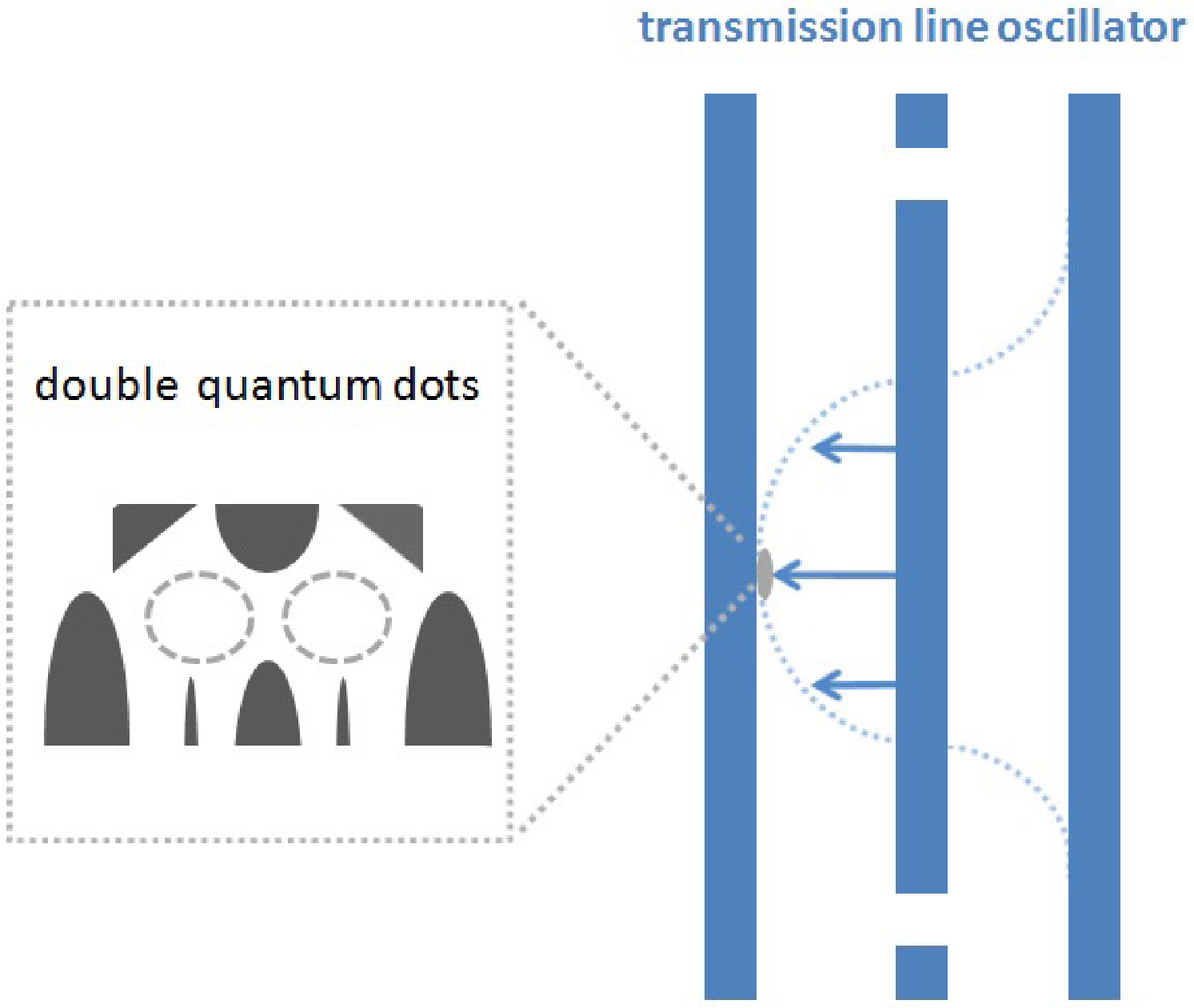}
                   \hspace{10mm}
\includegraphics[width=0.35\textwidth]{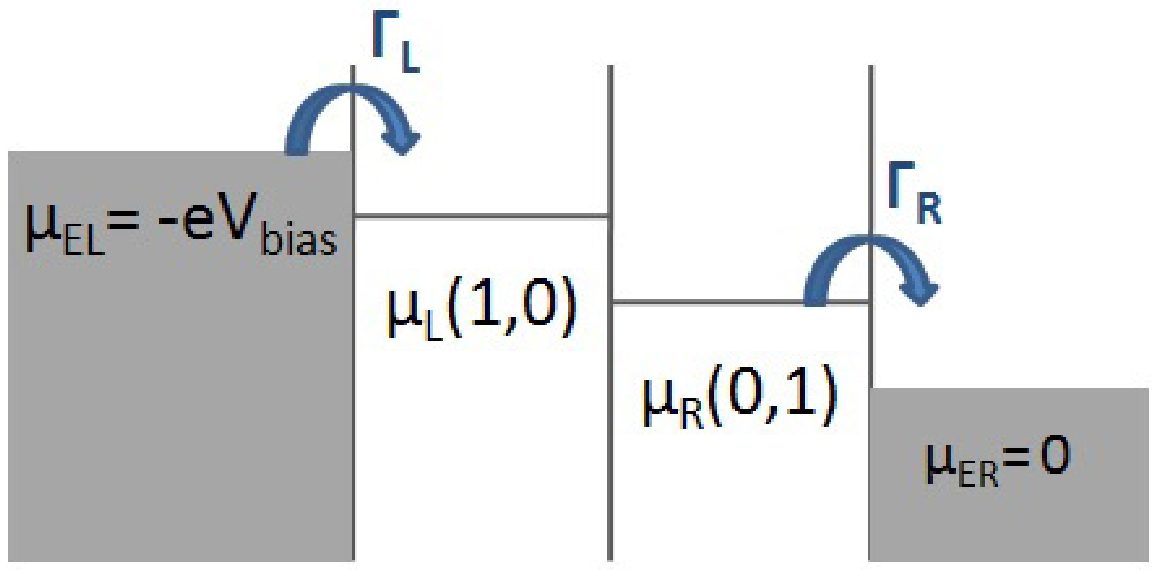}
\caption{Left panel: Illustration of a double quantum dot-resonator
circuit. The dot is placed at a maximum of the electric field of a coplanar waveguide (CPW).
Right panel: Tunneling sequence in the double dot system.
The electro-chemical potentials $\mu_{\rm L} (1,0)$ and $\mu_{\rm R} (0,1)$ of the two dots
are assumed to be arranged as indicated.  }\label{fig:dot}
\end{figure}

Here we propose a different CQED setup where a double quantum dot is
coupled to a superconducting resonator in a geometry indicated in
Fig. \ref{fig:dot}.
The double quantum dot is biased such that a single electron occupies the
lowest empty orbital in the left or right dot. These two relevant states
are denoted as pure charge states $|1,0\rangle$ and $|0,1\rangle$
with an energy difference $\epsilon$. A coherent interdot tunneling
with strength $t$ is assumed to couple the two
pure charge states. The resulting eigenstates are mixtures of pure
charge states with mixing angle $\theta = \arctan(t/\epsilon)$.
Transitions between these eigenstates are allowed due to an electrical
dipole interaction with the resonator. Under the rotating wave
approximation, the Hamiltonian for the coupled system in the
eigenbasis of the two-level system is given by the standard
Jaynes-Cumming Hamiltonian
\begin{eqnarray}\label{eq:ChargeH}
 H_{\rm{sys}}  = \frac{\hbar\omega_0}{2} \sigma_z
 + \hbar \omega^{}_{\rm r} a^\dag a
 + \hbar g (a^\dag \sigma_- +a^{} \sigma_+).
\end{eqnarray}
Here $\omega_0 = \sqrt{\epsilon^2+t^2}/\hbar$ and $\omega_{\rm r}$
denote the frequencies of the two-level system and the resonator,
respectively, where a detuning $\Delta = \omega_0-\omega_{\rm r}$ is
allowed, and $a^{}$ ($a^\dag$) represents the annihilation
(creation) operator of photons in the resonator.

The dynamics of the coupled dot-resonator system is studied in the
frame of a master equation for the reduced density matrix $\rho$.
Within the conventional Born-Markovian approximation, the master
equation is given by
\begin{eqnarray} \label{eq:ME}
 \dot \rho &=& -\frac{i}{\hbar}\left[H_{\rm_{sys}}, \rho\right]
  + \sum_i \frac{\Gamma_i}{2}\left(2L_i\rho L_i^{\dag}-L_i^{\dag}L_i\rho-\rho
  L_i^{\dag}L_i\right).
\end{eqnarray}
The dissipative dynamics is described in Lindblad form with operators
$L_i$. For the decay of resonator, we adopt the standard form
$L_{\rm r}=a$ with rate $\Gamma_{\rm r}=\kappa$. For relaxation and
decoherence of the two-level system, the corresponding Lindblad
operators are given by $L_\downarrow =\sigma_-$ with rate
$\Gamma_\downarrow$ and $L_{\varphi}=\sigma_z$ with rate
$\Gamma_\varphi^*$. Throughout the paper we consider low temperatures
with vanishing thermal photon number and excitation rates. The incoherent tunneling between electrodes and
quantum dots are also included (see below).

To achieve the population inversion for lasing we assume the system
to be biased such that only the electro-chemical potentials of the states
$|1,0\rangle$ and $|0,1\rangle$ lie within the bias window. At low
temperatures compared to the charging energy, an electron can only
tunnel into the dot system from the left electrode to the left dot.
It leads to a transition from the state $|0,0\rangle$ to
$|1,0\rangle$. This process is described by a Lindblad operator $L_{\rm
L}=|1,0\rangle\langle 0,0|$. Similarly, an electron can tunnel out
into the right lead, creating a transition from $|0,1\rangle$ to
$|0,0\rangle$, which is described by
$L_{\rm R}=|0,0\rangle\langle 0,1|$. For simplicity, we assume these
two processes having the same tunneling rate $\Gamma$. A
non-equilibrium state is achieved with enhanced population in the
state $|1,0\rangle$ with higher energy as compared to the state
$|0,1\rangle$. This effect persists in the eigenbasis. The resulting
population inversion in the absence of relaxation between dot
levels, becomes
\begin{eqnarray}
\tau_0 =\frac{4\cos\theta}{3+\cos(2\theta)}.
\end{eqnarray}
As the system approaches the degeneracy point with $\theta = \pi/2$,
the pure charge states are coupled strongly by the interdot
tunneling, and the population inversion decreases.

\begin{figure}[t]
\centering
\includegraphics[width=0.38\textwidth]{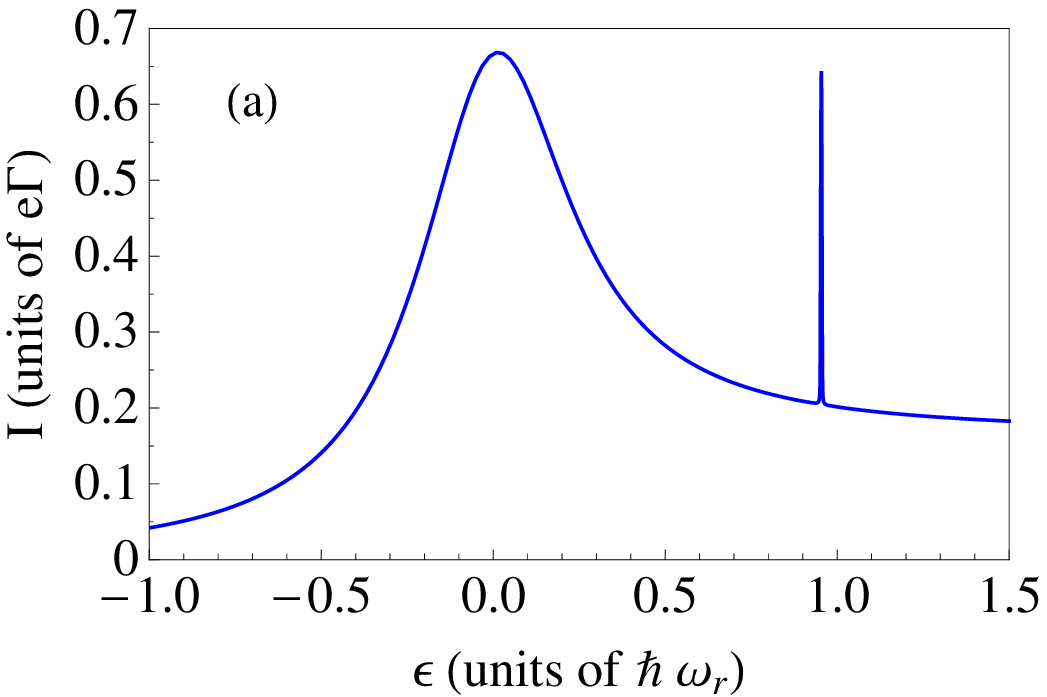}
 \hspace{5mm}
\includegraphics[width=0.40\textwidth]{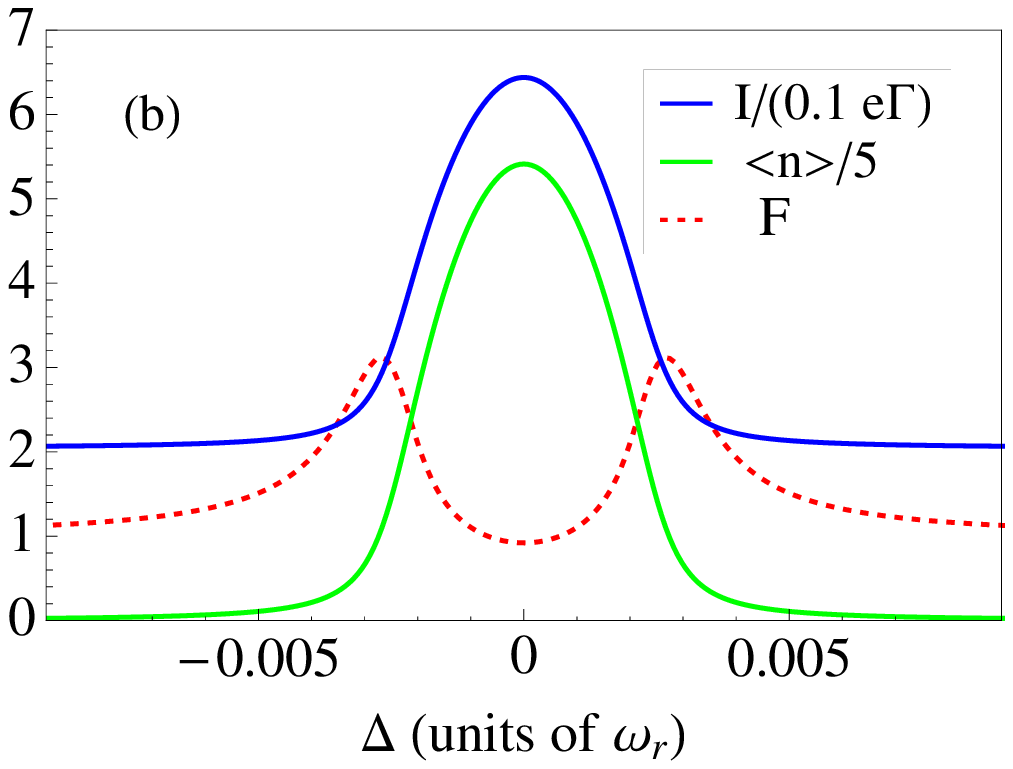}
\caption{(a) Current as a function of the energy
difference $\epsilon$. (b) Average photon number $\langle n\rangle$,
Fano factor $F$ and transport current $I$ as functions of the
detuning. In both panels, we choose the interdot tunneling strength
$t = 0.3~\hbar\omega_{\rm r}$, incoherent tunneling rate $\Gamma
=10^{-3}\omega_{\rm r}$, decay rate $\kappa = 10^{-5} \omega_{\rm
r}$, relaxation and decoherence rates $\Gamma_\downarrow = \Gamma_\varphi^* = 10^{-4} \omega_{\rm r}$,
and coupling strength $g = 3\times 10^{-4}\omega_{\rm r}$.
 }\label{fig:Det}
\end{figure}

A current flows through the double dot when the tunneling cycle
$|0,0\rangle\rightarrow|1,0\rangle\rightarrow|0,1\rangle\rightarrow|0,0\rangle$
is completed. We evaluate the current using $I =
e\sum_{i,j}\Gamma_{i\rightarrow j}\,\langle i|\rho_{\rm st}
|i\rangle$, where the index $i$ refers to the eigenstates of the
two-level system as well as the state $|0,0\rangle$, and
$\Gamma_{i\rightarrow j}$ denotes the transition rate from state
$|i\rangle$ to $|j\rangle$. The transport current at low temperature
(vanishing thermal photons) is plotted in Fig.~\ref{fig:Det} as a function of the energy difference $\epsilon$.
At the degeneracy point $\epsilon = 0$, a broad peak shows up
due to the coherent interdot tunneling, which width is given by the tunneling
rate $t$ (here $t\gg \hbar\,\Gamma$)~\cite{Vaart,Stoof}.
With respect to the degeneracy point, the current exhibits asymmetric behavior.
On the absorption side ($\epsilon <0 $), an electron that tunnels into the left dot is blocked
in the state $|1,0\rangle$ since no photon is present to provide the energy for tunneling
to the right dot. Therefore on the absorption side, only the residue from the broad current peak remains.
On the emission side ($\epsilon >0 $), a second current peak arises
when the two-level system becomes resonant with the oscillator.
For realistic values of the parameters, this current peak is much sharper.
It correlates with a lasing state within a narrow resonance window
with width $W\approx 2 g\sqrt{\Gamma/\kappa-1}$ for small
$\theta$ \cite{DotResonator}, since a photon is generated in the resonator when an electron
tunnels between the two dots.
Besides, the relaxation of the two-level system, which opens up an incoherent channel,
increases the overall current on the emission side.

The lasing state is characterized by the average photon
number $\langle n\rangle$ as well as the Fano factor $F \equiv
(\langle n^2\rangle-\langle n\rangle^2)/\langle n\rangle$.
As indicated in Fig. \ref{fig:Det}, at low temperature, the photon number vanishes for large detuning,
since the quantum dot system does not interact effectively with the
resonator. The system is in the non-lasing regime and the Fano factor can be approximated by $F\simeq\langle
n\rangle +1$. When moving towards the resonance,
the system undergoes a lasing transition, accompanied
by a sharp increase in the photon number, and an enhanced Fano
factor since the amplitude fluctuations increase. At resonance, the
photon number reaches a maximum, and the Fano factor is near 1,
indicating the system is in a coherent state.

To study experimentally the predicted lasing effect, it is most suitable to process a double-quantum-dot system with in-situ tunable parameters which allow controlling the electron numbers and energy levels within the quantum dots but also the tunnel couplings towards the leads and/or between the quantum dots. A double quantum dot system based on a 2D electron system (2DES) which is embedded in a GaAs/(AlGa)As heterostructure close to the heterostructure surface offers such properties.
The 2DES is structured by either applying negative voltages to structured metal electrodes on top of the heterostructure surface, or by etching grooves into the heterostructure to the depth of the 2DES and thereby dividing the 2DES into regions acting as leads, gates, and quantum dots. The second approach is more suitable in our case as the CPW resonator can be directly deposited on the heterostructure without being affected by normal conducting metal electrodes which might cause energy loss for the resonator. By removing the 2DES in most areas of the resonators's mode volume -- except for the region of the quantum dot system in the center of maximum field of the resonator -- the interaction of the electrically dissipative 2DES with the resonator field is diminished.

In such an approach, the typical geometrical diameter of a quantum dot is 0.5 to 0.7 $\mu$m, leading -- due to electrostatic depletion from the etched grooves -- to an effective  quantum dot diameter of about 0.2 to 0.4 $\mu$m. The respective
single-electron charging energies ($e^2/2C$) are then in the range of 0.5 to 1 meV.
Excitations in such quantum dots can be characterized by electrical transport spectroscopy and are found at energies which are typically 1/10 to 1/3 of the single-electron charging energy. A disadvantage of such structures might be that the tunable tunnel barriers are energetically shallow and spatially broad.

The proposal here relies on a double-quantum dot where a single electron jumps from one dot to the other. The respective energy difference $\epsilon$ for the electron sitting on either quantum dot is tunable by applying gate voltages. The respective electrical dipole change $d$ which couples to the resonator field is about the distance between both quantum dot centers times the elementray charge, i.e. here about $d=0.6\,\mu\textrm{m}\times e$. For an electric field strength of $E=0.2\,\textrm{V}\cdot\textrm{m}^{-1}$ - typical of such a CPW resonator, we find a maximum coupling factor of $g=d\cdot E/\hbar\approx 20\,\textrm{MHz}$.

Achieving the lasing action puts constraints on parameters, namely, a relatively strong coupling compared to dissipations \cite{SQL2},
\begin{equation}
g > \sqrt{\frac{\kappa\Gamma_{\varphi} }{2\tau_0}},
\end{equation}
where $\Gamma_\varphi$ denotes the total rate accounting for all sources of dissipation of the two-level system.
For small $\theta$ it reduces to $\Gamma_\varphi \simeq \Gamma_\downarrow/2 + \Gamma_\varphi^* + \Gamma/4$.
To present date, CPW resonators on GaAs substrates with $\kappa=10^{-4}\omega_{\textrm r}$ have been reported~\cite{frey2011},  two orders of magnitude larger than achieved on other substrates. This value of $\kappa$ would be sufficient to obtain lasing, if  $\Gamma_{\varphi}$  can be kept in the range of 10 MHz.

On the other hand, the lifetime of an electron inside the double-quantum dot system has to be long enough to allow for stimulated photon emission while making the transition from one dot to the other. The respective tunnel rates can be tuned to low value. However, we want to detect currents,  which under optimum conditions requires current above 1 pA, i.e. the tunneling rates have a lower limit given by $1/\tau < I/e \approx 0.6\,\textrm{MHz}$. This would suffice to detect the lasing peak of magnitude about $2e\Gamma/3$~\cite{DotResonator}. Given sufficiently high coupling it should also be possible~\cite{schuster2007} to observe the change in photon number characteristic of lasing.

\ack

We acknowledge helpful discussion with S. Andr\'e and A. Romito, as
well as the financial support from the Baden-W\"{u}rttemberg
Stiftung via the Kompetenznetz Funktionelle Nanostrukturen.

\end{document}